\documentclass[twocolumn,prl,aps,epsf, reprint footinbib]{revtex4-1}
\usepackage{graphicx}
\usepackage{dcolumn}
\usepackage{bm}
\usepackage{hyperref}
\usepackage{mathtools}

\begin{document}

\title{Specific heat and entropy of fractional quantum Hall states in the second Landau level}

\author{B. A. Schmidt$^{1}$, K. Bennaceur$^{1}$, S. Gaucher$^{1}$,  G. Gervais$^{*1}$, L. N. Pfeiffer$^{2}$ and K. W. West$^{2}$}

\affiliation{$^{1}$ Department of Physics, McGill University, Montreal, H3A 2T8, CANADA}

\affiliation{$^{2}$ Department of Electrical Engineering, Princeton University, Princeton NJ 08544 USA}

\date{\today }

\begin{abstract}
Specific heat has had an important role in the study of superfluidity and superconductivity, and could provide important information about the fractional quantum Hall effect as well.  However, traditional measurements of the specific heat of a two-dimensional electron gas are difficult due to the large background contribution of the phonon bath, even at very low temperatures. Here, we report measurements of the specific heat per electron in the second Landau level by measuring the thermalization time between the electrons and phonons. We observe activated behaviour of the specific heat of the 5/2 and 7/3 fractional quantum Hall states, and extract the entropy by integrating over temperature. Our results are in excellent agreement with previous measurements of the entropy via longitudinal thermopower. Extending the technique to lower temperatures could lead to detection of the non-Abelian entropy predicted for bulk quasiparticles at  $5/2$ filling.
 \end{abstract}

\maketitle 

\textit{Introduction} --- An ultra clean two dimensional electron gas (2DEG) exposed to high magnetic fields and low temperatures plays host to a rich array of phases, including the intrinsically topological fractional quantum Hall (FQH) states. The $\nu = 5/2$ FQH state is of particular interest, since it is believed to obey non-Abelian statistics \cite{Willett1987, Moore1991}. Unfortunately, fabrication of devices to study the FQH states, such as quantum point contacts and interferometers, often degrades the quality of the sample, rendering the 5/2 FQH effect unobservable. In cases where studies have been performed, their interpretation is difficult due to our incomplete understanding of the detailed physics of the quantum Hall edge. In this paper, we introduce a technique to probe the bulk of the 5/2 FQHE, avoiding the edge entirely. In particular, we report measurements of the specific heat at $\nu = 5/2$, which, unlike transport, is sensitive to the total density of states (DOS). Moreover, one of the signatures of a non-Abelian system is an excess ground state entropy $S_{NA} = k_B N_{qp} \ln d$, where $N_{qp}$ is the number of quasiparticles and $d$ is the quantum dimension (equal to $\sqrt{2}$ for the conjectured non-Abelian Pfaffian and anti-Pfaffian states at 5/2) \cite{Cooper2009, Yang2009}. In principle, this entropy could be detectable in the specific heat in the low temperature limit \cite{Cooper2009}.

Measurement of the specific heat of a 2DEG within a heterostructure is difficult because it is dwarfed by the contribution of the substrate. Earlier studies \cite{Gornik1985, Wang1988, Bayot1996} applied conventional measurement techniques to multiple quantum well structures to boost the relative contribution of the electronic signal. In a more recent \emph{tour de force} study, Schulze-Wischeler \emph{et. al.} \cite{Schulze-Wischeler2007} applied a phonon absorption technique to determine the specific heat in the FQH regime, albeit in arbitrary units. Our experiment similarly uses the weak electron-phonon coupling at low temperature to thermally isolate the 2DEG (on sufficiently short timescales), however we make use of \emph{in situ} Joule heating and extract an absolute value for the specific heat.  Furthermore, we use a Corbino disk, in which no edges connect the inner and outer contacts and we can be certain that we are probing the bulk of the 2DEG \cite{Barlas2012}. The radial symmetry of the Corbino geometry also simplifies analysis, since we can neglect the Nernst, Ettingshausen and thermal Hall effects, which can strongly affect the temperature distribution in Hall and van der Pauw samples \cite{Hirayama2011}. \\

\textit{Experimental Overview} --- Our experimental protocol consists of three conceptual parts, which are performed in an interlaced fashion in order to minimize effects of drift. The first is to measure the conductance of the sample, $G$, as a function of electron temperature, $T_e$, using an excitation small enough that $T_e$ remains close to the refrigerator temperature, $T_0$. At certain filling factors and temperatures, we find that $G$ is highly temperature sensitive and thus can act as a thermometer for the 2DEG. Next, we apply a singled-sided square wave at several kHz with a large (typically millivolt-scale) bias $V_{high}$ which both heats the electrons and allows us to measure the 2DEG conductance, as shown in Fig~1B.  The thermal time constant, $\tau$, can then be extracted from an exponential fit to the conductance transient response as shown in Fig~1C. This may be done either for the turn-on portion, as presented in the main body of the letter, or as the electrons cool down again (using a small, but non-zero bias for $V_{low}$), as discussed in the supplemental material \cite{Supp}. Finally, we apply a series of DC biases to the sample, while measuring its conductance. Using the calibration of $G (T_e)$, we deduce the electron temperature as a function of applied DC power and phonon temperature. From this we extract $K$, the total thermal conductance between the electronic system and the environment. The total heat of the system can then be found from the relation $C = K \tau$  \cite{Bachmann1972}.\\

\begin{figure}
  \centering
      \includegraphics[width=\columnwidth]{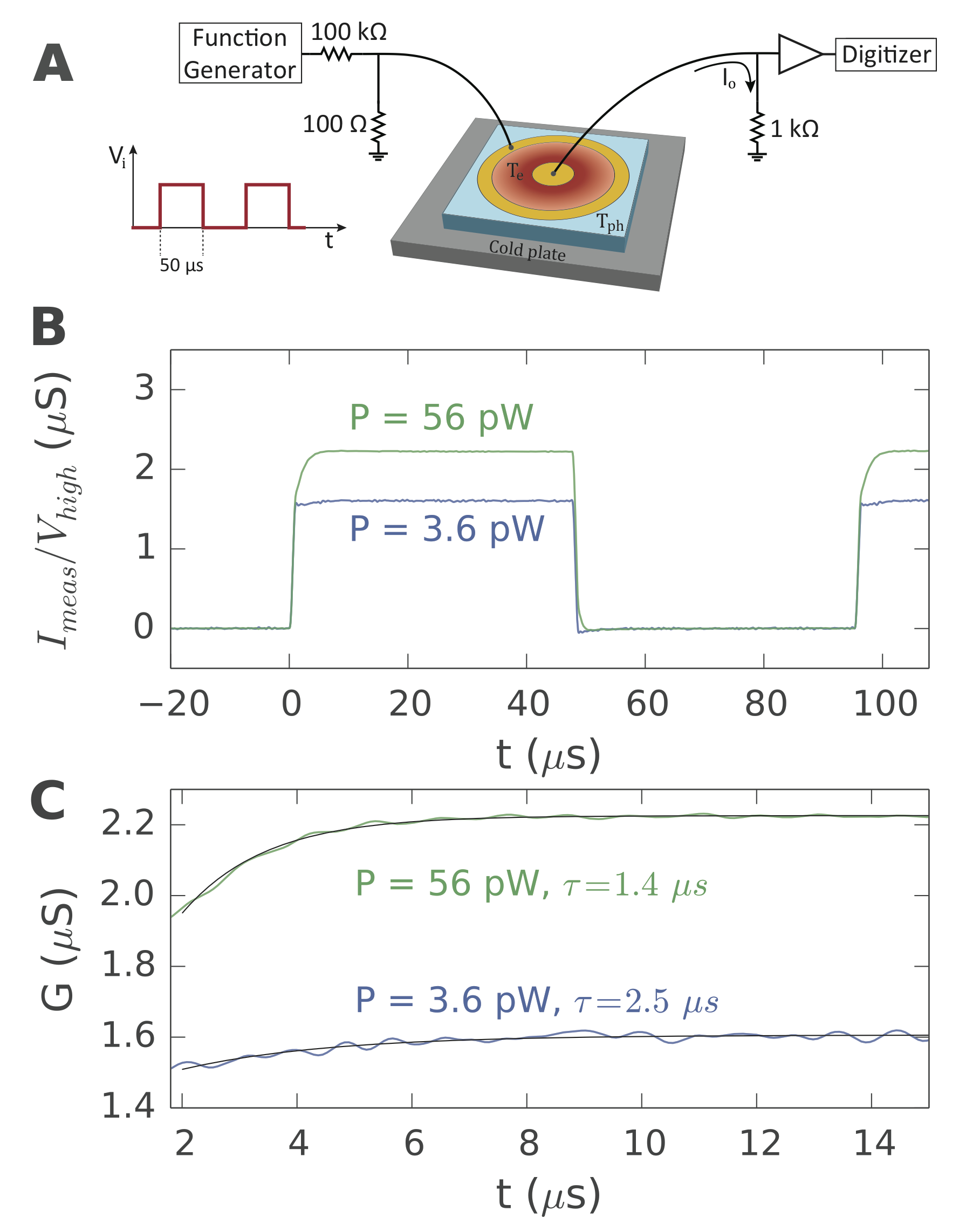}
  \caption{(\textbf{A}) Experimental scheme. A square wave voltage bias is applied between the gold contacts (yellow) to heat the 2DEG (red) above the lattice temperature of the GaAs wafer (blue). The resulting current is measured across a 1~k$\Omega$ resistor to ground, using a voltage preamplifier and a digitizer. (\textbf{B}) Example time traces showing the measured current response. Each trace was obtained by averaging approximately $10^6$ iterations. (\textbf{C}) Zoomed in plot of the conductance transient. Exponential fits are shown by the solid black lines, and corresponding values of $\tau$ are given for each curve. }
\end{figure}

\textit{Experimental details} --- All measurements presented in this letter were performed in a GaAs/AlGaAs heterostructure with quantum well width 30~nm, electron density $n_e \approx 3.06 \times 10^{11} \,{\rm cm^{-2}}$ and wafer mobility ${\rm 2.5 \times 10^{7} cm^2/V \cdot s}$ measured at 0.3~K. The Corbino device was defined by a central contact with outer radius $r_1 = 0.25$~mm and a ring contact with inner radius $r_2 = 1.0$~mm. Full fabrication and characterization details can be found in reference \cite{Schmidt2015}.

Values of $K$ are extracted from square wave response data. First, $G (T_{0}, P)$ is determined from the average of $I_{meas}/V_{high}$ \textit{after} the thermal transient - for example, between $t = 15~{\rm \mu s}$ and $t = 47~{\rm \mu s}$ in Fig.~1B. Then, a smooth cubic spline interpolation is fit to $T_{ph}$ vs. $G$ in the low power limit, where $T_e \simeq T_{0}$, and used to find  $T_e (T_{0}, P)$ for higher heating powers. Finally, we calculate $K$ from the slope of $T_e$ vs $P$  for $P$ small enough to only raise the electron temperature by a few millikelvin. Further experimental details, including a correction factor for the Corbino geometry, are provided in the supplemental material \cite{Supp}.

The same data set used to find $K$ is also used to find $\tau$. As shown in Fig.~1B, a transient is seen after the bias is turned on, but not when the voltage is turned off (since there is no voltage to convert the conductance into a current). 
Figure~1C shows an expanded view of the transient, after correction for possible LRC transients \cite{Supp}.
The fitted time constant is shorter at higher power, since the 2DEG reaches a higher temperature and therefore has a higher energy emission rate. In order to find $\tau (T_e)$, we associate each measured time constant to the electron temperature inferred from the final conductance it reaches after several microseconds.

\begin{figure}
  \centering
      \includegraphics[width=\columnwidth]{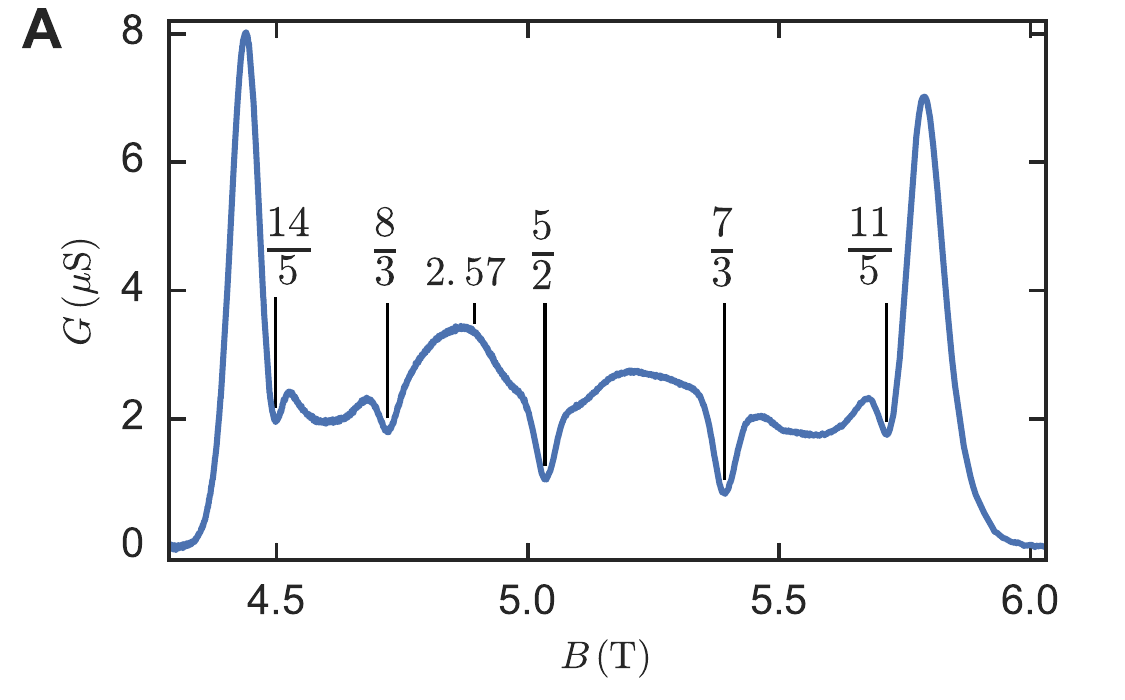}\\
      \includegraphics[width=\columnwidth]{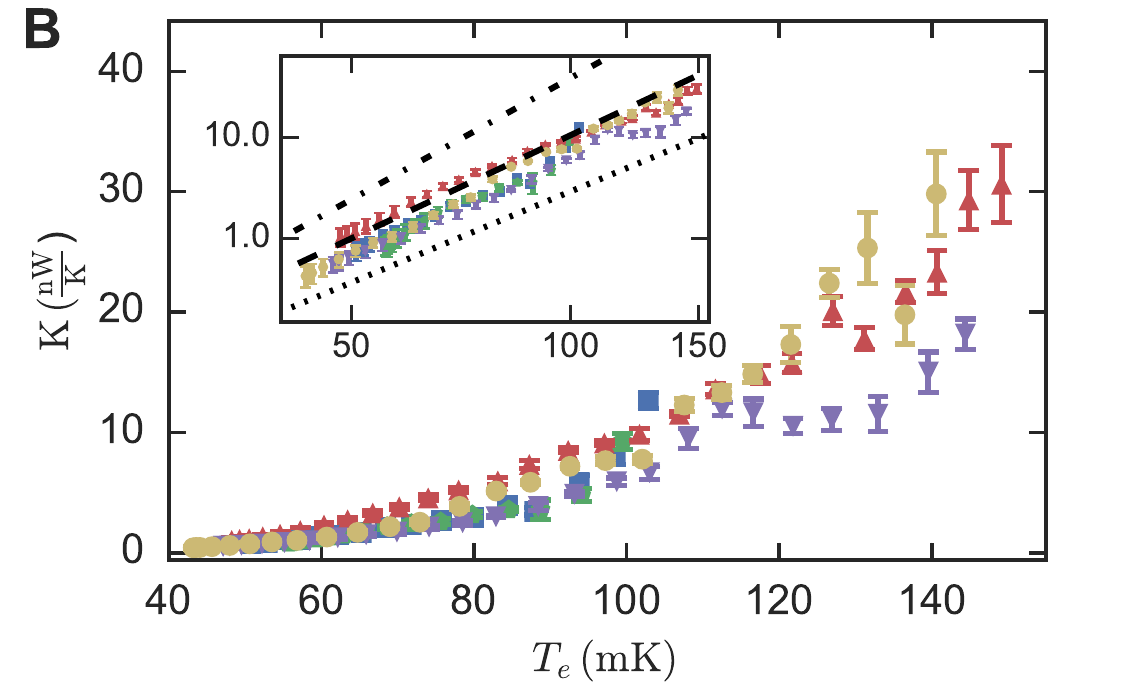}
      \includegraphics[width=\columnwidth]{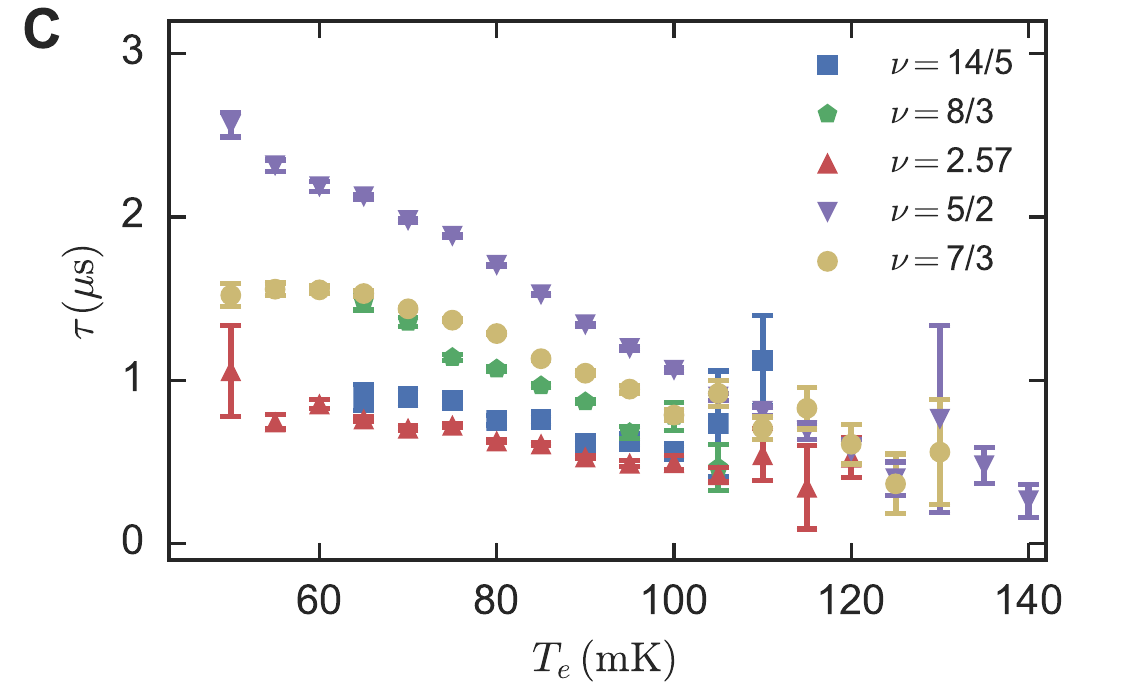}\\
  \caption{(\textbf{A}) Conductance at base temperature. Arrows indicate filling factors where measurements of $\tau$, $K$ and $C$ were performed. (\textbf{B}) $K$ vs $T_e$ for several filling factors in the SLL. Legend is provided in panel (C). The inset shows the same data on a log-log scale with lines, at arbitrary vertical positions, indicating slopes of 3 (dotted), 3.4 (dashed) and 4 (dot-dashed). (\textbf{C}) Thermal relaxation time $\tau$ measured as a function of $T_e$ (as calculated from conductance measurements). Data from multiple phonon temperatures have been binned based on $T_e$ in 5~mK bins and averaged. }
\end{figure}

\textit{Results}--- We focus on several filling factors in the SLL that are marked on Fig.~2A, which shows the sample's conductance at base temperature. Most of these filling factors are weakly-gapped FQH states, however we also measured at $\nu = 2.57$ where we observed a markedly increasing conductance with decreasing temperature. At lower temperatures, a reentrant integer quantum Hall state, is often observed 
\cite{Eisenstein2002, Deng2011} at that same filling factor. Select results in high filling factors ($\nu > 10$) are presented in the supplemental material \cite{Supp}, with details to be presented in a separate publication. 

\textit{Thermal conductance to the environment}--- Our results for $K$ are shown in Fig.~2B. The trend is similar for all of the filling factors shown, although ${\nu=2.57}$ exhibits a somewhat higher value of $K$ throughout the temperature range below $100$~mK. The magnitude of $K$ is several orders of magnitude larger than would be expected according to Wiedemann-Franz law ($K = 12 G L_0T \approx 10 $~fW/K, where $L_0$ is the Lorenz number and the factor of 12 arises from geometric considerations), effectively ruling out diffusion of charged quasiparticles to the contacts as a dominant cooling mechanism. Within the temperature range of our experiment, our results are instead consistent with cooling by phonon emission, although in principle we cannot rule out thermal transport by neutral quasiparticles \cite{Bonderson2011a}.

At $B=0$, the problem of electron-phonon power emission has been studied theoretically by Price and others \cite{Price1982, Karpus1988}, and good experimental agreement was found by Appleyard \textit{et. al.} \cite{Appleyard1998}. Using that model, we would expect $K = 370\; T^4$~[nW/K] for our sample geometry, which is roughly two orders of magnitude lower than what we observe. However, this is consistent with the enhancement of CF-phonon scattering (relative to electron-phonon scattering at $B=0$) seen in experiments measuring power emission power\cite{Chow1996}, phonon-drag thermopower \cite{Tieke1998} and phonon-limited mobility \cite{Kang1995}, as well as a theoretical treatment of CF-phonon interactions \cite{Khveshchenko1997}. Fitting $K \propto T^n$, we find $n \simeq 3.4$ (indicated by the dashed black line in figure~2c), which is between the value $n=4$ in the model by Price \cite{Price1982} and $n=3$ in the hydrodynamic model put forward by Chow \emph{et. al.} \cite{Chow1996}. Both of those models used a flat (metallic) DOS for the 2D electrons, and would have to be modified to take into account the gapped DOS at FQH states.

\textit{Thermal relaxation time} --- Figure~2b shows $\tau$ as a function of electron temperature for each state.  These are, to our knowledge, the first direct measurements of electron-phonon energy relaxation times in the quantum Hall regime below 100~mK. Previous estimates of a $T^4$ or $T^3$ dependence for $\tau$ were based on measurements of DC electron-phonon energy emission rates and assumed linear behaviour of $C(T)$ as calculated for a 2DEG at zero field \cite{Gammel1988, Mittal1996, Chow1996}. Since we are measuring at (or near) gapped states, we do not expect $C(T)$ to be linear and we do not attempt to fit $\tau(T)$ with simple power law. While $\tau$ is monotonically decreasing in all cases, there are clear differences between filling factors. The thermal relaxation time at $\nu = 5/2$ is slower than those at $\nu = 7/3$ and $\nu = 8/3$, which are in turn slower than at $\nu=2.57$ and  $\nu=14/5$. The apparent differences in $\tau$ may be due to differences in the charge, size and screening of quasiparticles at each filling factor. 

\begin{figure*}
  \centering
      \includegraphics[width=1.0\textwidth]{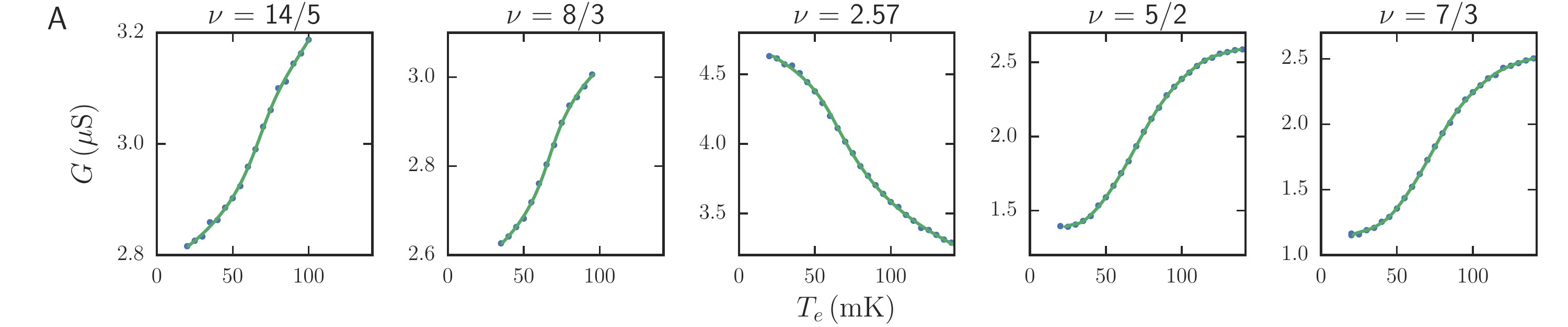}
      \includegraphics[width=1.0\textwidth]{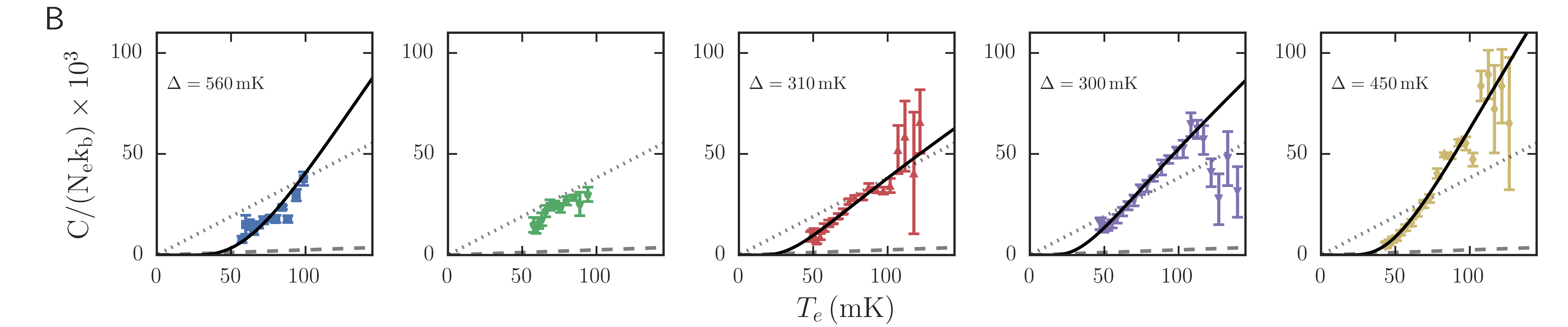}
  \caption{(\textbf{A}) Conductance vs. temperature for several states in the SLL.   
  (\textbf{B}) Specific heat vs. temperature at the same set of filling factors. 
 Fits to equation~\ref{eqn:C_gap} are shown by the solid black lines, and the resulting $\Delta$'s are given on each plot,  while $g_0k_B/n_e$ = 0.18, 0.12, 0.16 and 0.25~K$^{-1}$ for $\nu =$~14/5, 2.57, 5/2 and 7/3, respectively. The dotted line is the specific heat for free 2D electrons ($m^* = m_e$), while the dashed line is the specific heat for 2D electrons in GaAs at zero field ($m^* = 0.067m_e$). The fit at $\nu = 14/5$ is of lower quality than the others, and is discussed further in \cite{Supp}.}
  \label{fig:gaps}
\end{figure*}

\textit{Specific heat} --- The electronic heat capacity is now given by $C = K \tau$.  Fig.~3C shows the calculated specific heat, $c \equiv C/k_BN_e$, where $N_e$ is the total number of electrons in the Corbino disk. For comparison, we also show conductance vs. temperature plots at each filling factor in Fig.~3A. We begin our analysis by considering the specific heat of a fermi liquid, given by

\begin{equation}
c = \frac{\pi m^* k_B T}{3 \hbar^2 n_{q}},
\end{equation}

where $m^*$ is the effective mass of the fermions and $n_q$ is the number of quasiparticles. Using the band mass of GaAs, $m*=0.067m_e$, and $n_q=n_e$, we obtain the specific heat at $B=0$ as shown by the dashed lines in Fig~3B. The observed specific heat is much larger - in fact, it agrees in magnitude with $c$ for a fermi liquid of free electrons ($m^* = m_e$, the dotted line in each panel of Fig~3B). This is in order-of-magnitude agreement with both theory \cite{Cooper1997} and experiments \cite{Chickering2013, Kukushkin2002} that have shown composite Fermions at half-filling to have an effective mass close to that of free electrons. However, the specific heat at each filling factor increases rapidly in the region from 50 to 100~mK, exhibiting a strong deviation from linear (fermi-liquid like) behaviour, as expected for a gapped DOS. For simplicity, we may consider the specific heat for a system of fermions with a toy DOS $g(\epsilon)$ given by two flat regions separated by a gap $\Delta$, with the fermi level exactly halfway between the two levels. From this model we obtain, in the low temperature limit,

\begin{equation}
c = \frac{2 k_Bg_0}{n_e} \left(\frac{\Delta^2}{4k_BT} + \Delta + 2k_BT\right)e^{-\Delta/2k_BT} ,
\label{eqn:C_gap}
\end{equation}

which is the equation used to fit the black curves in Fig.~3B. In principle, the effective mass of the quasiparticles is given by  $m^* = \pi \hbar^2 g_0$, and is found be $1.4m_e$ at 5/2 and $2.1m_e$ at 7/3. However, such analysis should only be taken as a rough estimate, since it is based on a simplified model of the DOS. Given the limited temperature range of our data set, a generic Arrhenius behaviour also fits well to the data and yields a similar value for the gap energy \cite{Supp}. Interestingly, standard Arrhenius fits to the conductance yield significantly smaller gap energies - specifically, $\Delta_{5/2} = 103 \, {\rm mK}$ and $\Delta_{7/3} = 131 \, {\rm mK}$  \cite{Supp}. The discrepancy can be understood by considering more detailed models of conductance, such as the saddle point model proposed by d'Ambrumenil \textit{et. al} \cite{Dambrumenil2011}. In their framework, the naive Arrhenius fit to conductance systematically underestimates the true energy gap. Using their recipe to estimate the true gap from conductance, we obtain $\Delta_{5/2} = 300 \, {\rm mK}$ and $\Delta_{7/3} = 330 \, {\rm mK}$, in good agreement with the results from specific heat.

\begin{figure}
  \centering
      \includegraphics[width=0.9\columnwidth]{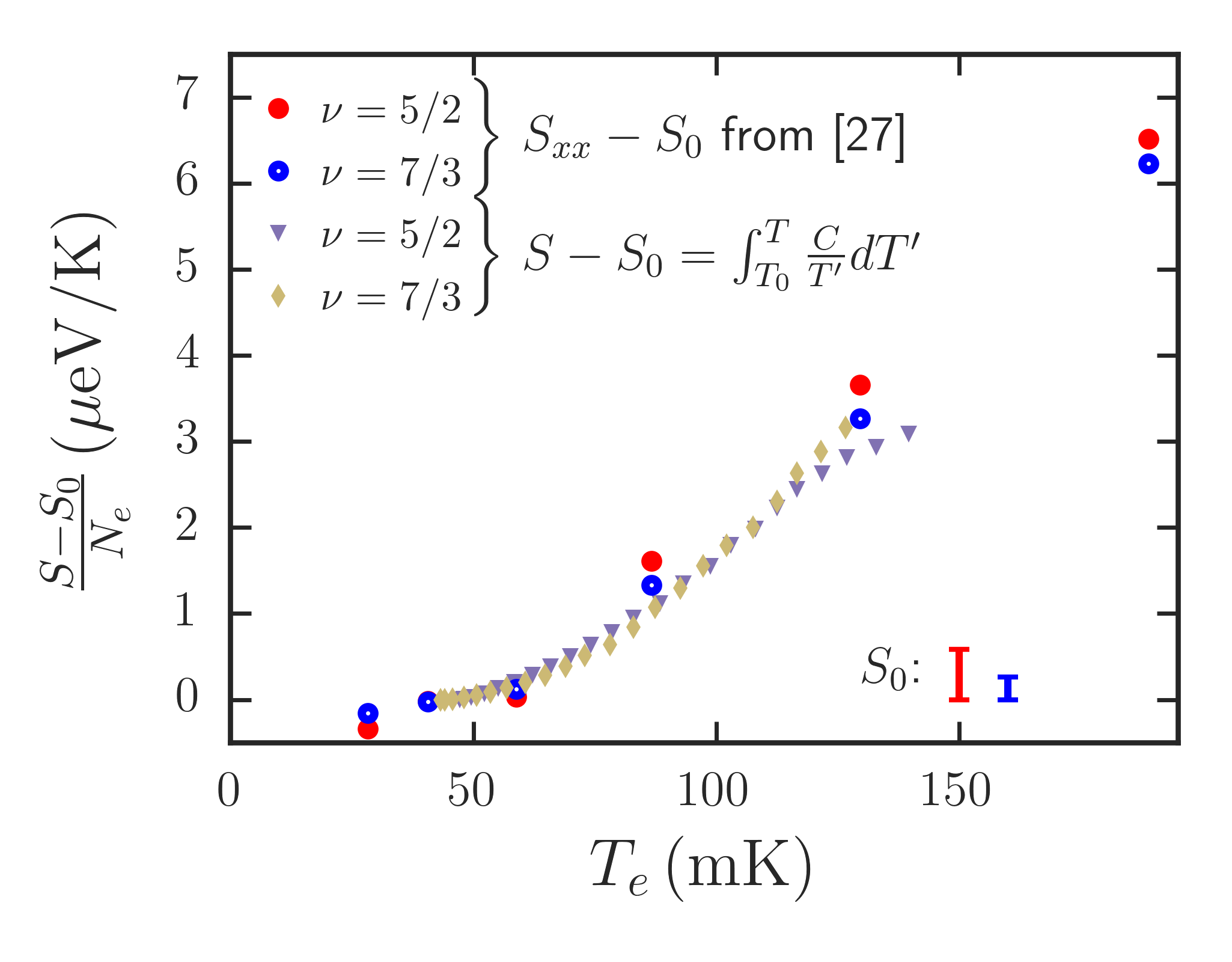}
  \caption{ Entropy as determined from longitudinal thermopower data \cite{Chickering2013} and from integration of our specific heat measurements. The thermopower data is offset such that $S-S_0=0$ at the lowest temperature for which we measured $C$. The offsets are $S_0=0.58$ and $S_0=0.26$ for $\nu=5/2$ and $\nu=7/3$, respectively, as shown by the scale bars on the figure.}
  \label{fig:gaps}
\end{figure}

\textit{Entropy} --- The entropy of a 2DEG can also be accessed by measuring the longitudinal thermopower, $S_{xx}$, which is related to entropy by the relation ${S_{xx}=-S/(|e|n_e)}$ in the clean limit \cite{Yang2009}. In order to compare our results to existing thermopower data, we extract the entropy as a function of temperature using the formula

\begin{equation}
S(T) - S_0 = \int^T_{T_0} \frac{C}{T'} dT',
\end{equation}

with $T_0$ being the lowest temperature at which we measured $C$. In figure~4, we plot $S(T) - S_0$, calculated from our measurements of $C$ by numerical integration along with thermopower data obtained by Chickering \textit{et. al.} (figure 2 of ref \cite{Chickering2013}). The data from these two completely different techniques are in excellent agreement, suggesting that both do indeed measure the entropy of the electron system in the SLL. 

The apparent activation-like behaviour at $\nu=2.57$ can also be understood by looking at longitudinal thermopower data. Chickering \textit{et. al.} observed a step in $S_{xx}$, corresponding to onset of the re-entrant state and superlinearly increasing $S_{xx}$ at higher temperature \cite{Chickering2013}. We do not observe the onset of the reentrant state itself, however, we do see superlinear (activation-like) behaviour of $c$, in qualitative agreement with the thermopower result.

\textit{Conclusion} ---  We have directly measured the electron-phonon energy relaxation rate and phonon emission power for several filling factors in the SLL. We observe clear variation in thermal relaxation times between filling factors, with $\nu=5/2$ in particular cooling more slowly than the other fractions in the SLL. We extract the specific heat for each filling factor, and find the expected activation behaviour. Our results quantitatively agree with the entropy inferred from thermopower data, consistent with both techniques independently measuring the entropy of the 2DEG in the SLL. Further measurements at lower temperatures could be used to search for the non-Abelian entropy, and perhaps identify the degeneracy temperature for non-Abelian anyons at 5/2.\\

\acknowledgments{
This work has been supported by NSERC, CIFAR and FRQNT. The work at Princeton University was funded by the Gordon and Betty Moore Foundation through the EPiQS initiative Grant GBMF4420, and by the National Science Foundation MRSEC Grant DMR-1420541. Sample fabrication was carried out at the McGill Nanotools Microfabrication facility. We gratefully thank K. Yang for useful discussions and P. F. Duc for assistance with preparation of the manuscript. Finally, we thank R. Talbot, R. Gagnon, and J. Smeros for technical assistance.}

\bibliographystyle{apsrev4-1}

\begin{thebibliography}{29}%

\bibitem{Willett1987} R. Willett, J. P. Eisenstein, H. L. St{\"{o}}rmer, D. C. Tsui, A. C. Gossard, and J. H. English, Phys. Rev. Lett. \textbf{59}, 1776 (1987).%
  
\bibitem{Moore1991} G. Moore and N. Read, Nucl. Phys. B \textbf{360}, 362 (1991).

\bibitem{Cooper2009} N. R. Cooper and A. Stern, Phys. Rev. Lett. \textbf{102}, 176807 (2009).
  
\bibitem{Yang2009} K. Yang and B. I. Halperin, Phys. Rev. B \textbf{79}, 115317 (2009).

\bibitem{Gornik1985} E. Gornik, R. Lassnig, G. Strasser, H. L. St{\"{o}}rmer, A. C. Gossard, and W. Wiegmann, Phys. Rev. Lett. \textbf{54}, 1820 (1985).

\bibitem{Wang1988}  J. K. Wang, J. H. Campbell, D. C. Tsui, and A. Y. Cho, Phys. Rev. B \textbf{38}, 6174 (1988).

\bibitem{Bayot1996} V. Bayot, E. Grivei, S. Melinte, M. Santos, and M. Shayegan, Phys. Rev. Lett.  \textbf{76}, 4584 (1996).%
  %
\bibitem{Schulze-Wischeler2007} F. Schulze-Wischeler, U. Zeitler, C. v. Zobeltitz, F. Hohls, D. Reuter, A. D. Wieck, H. Frahm, and R. J. Haug, Phys. Rev. B  \textbf{76}, 153311 (2007).%
  %
\bibitem{Barlas2012} Y. Barlas and K. Yang, Phys. Rev. B  \textbf{85}, 195107 (2012).%
%
\bibitem{Hirayama2011} N. Hirayama, A. Endo, K. Fujita, Y. Hasegawa, N. Hatano, H. Nakamura, R. Shirasaki, and K. Yonemitsu, Journal of Electronic Materials  \textbf{40}, 529 (2011).%

\bibitem{Supp} See Supplemental Material at [URL will be inserted by publisher] for additional experimental details and analysis. %
  
\bibitem{Bachmann1972} R. Bachmann, F. J. DiSalvo, T. H. Geballe, R. L. Greene, R. E. Howard, C. N. King, H. C. Kirsch, K. N. Lee, R. E. Schwall, H.-U. Thomas, and R. B. Zubeck, Rev. Sci. Instrum.  \textbf{43}, 205 (1972).%

\bibitem{Schmidt2015} B. A. Schmidt, K. Bennaceur, S. Bilodeau, G. Gervais, L. N. Pfeiffer, and K. W. West, Solid State Commun.  \textbf{217}, 1 (2015).%

\bibitem{Eisenstein2002} J. P. Eisenstein, K. B. Cooper, L. N. Pfeiffer, and K. W. West, Phys. Rev. Lett.  \textbf{88}, 076801 (2002).%

\bibitem{Deng2011} N. Deng, A. Kumar, M. J. Manfra, L. N. Pfeiffer, K. W. West, and G. A. Cs{\'{a}}thy, Phys. Rev. Lett. \textbf{108}, 086803 (2012).%

\bibitem{Bonderson2011a} P. Bonderson, A. E. Feiguin, and C. Nayak, Phys. Rev. Lett. \textbf{106}, 186802 (2011).%

\bibitem{Price1982} P. J. Price, J. Appl. Phys. \textbf{53}, 6863 (1982).%

\bibitem{Karpus1988} V. Karpus, Soviet Physics - Semiconductors \textbf{22}, 268 (1988).%

\bibitem{Appleyard1998} N. J. Appleyard, J. T. Nicholls, M. Y. Simmons, W. R. Tribe, and M. Pepper, Phys. Rev. Lett. \textbf{81}, 3491 (1998).%

\bibitem{Chow1996} E. Chow, H. P. Wei, S. M. Girvin, and M. Shayegan, Phys. Rev. Lett. \textbf{77}, 1143 (1996).%

\bibitem{Tieke1998} B. Tieke, R. Fletcher, U. Zeitler, M. Henini, and J. Maan, Phys. Rev. B \textbf{58}, 2017 (1998).%

\bibitem{Kang1995} W. Kang, S. He, H. L. Stormer, L. N. Pfeiffer, K. W. Baldwin, and K. W. West, Phys. Rev. Lett. \textbf{75}, 4106 (1995).%

\bibitem{Khveshchenko1997} D. V. Khveshchenko and M. Y. Reizer, Phys. Rev. Lett. \textbf{78}, 3531 (1997).%

\bibitem{Gammel1988} P. L. Gammel, D. J. Bishop, J. P. Eisenstein, J. H. English, A. C. Gossard, R. Ruel, and H. L. Stormer, Phys. Rev. B \textbf{38}, 10128 (1988).%

\bibitem{Mittal1996} A. Mittal, R. Wheeler, M. Keller, D. Prober, and R. Sacks, Surface Science \textbf{361-362}, 537 (1996).%

\bibitem{Cooper1997} N. R. Cooper, B. I. Halperin, and I. M. Ruzin, Phys. Rev. B \textbf{55}, 2344 (1997).%

\bibitem{Chickering2013} W. E. Chickering, J. P. Eisenstein, L. N. Pfeiffer, and K. W. West, Phys. Rev. B \textbf{87}, 075302 (2013).%

\bibitem{Kukushkin2002} I. V. Kukushkin, J. H. Smet, K. von Klitzing, and W. Wegscheider, Nature \textbf{415}, 409 (2002).%

\bibitem{Dambrumenil2011} N. D'Ambrumenil, B. I. Halperin, and R. H. Morf, Phys. Rev. Lett. \textbf{106}, 126804 (2011).%

\end{thebibliography}

\newpage

\renewcommand{\thefigure}{S\arabic{figure}}%
\setcounter{figure}{0}

\section{Supplementary Information}
\subsection {Subtraction procedure to isolate conductance transient}
\begin{figure}[!hbt]
  \centering
      \includegraphics[width=\columnwidth]{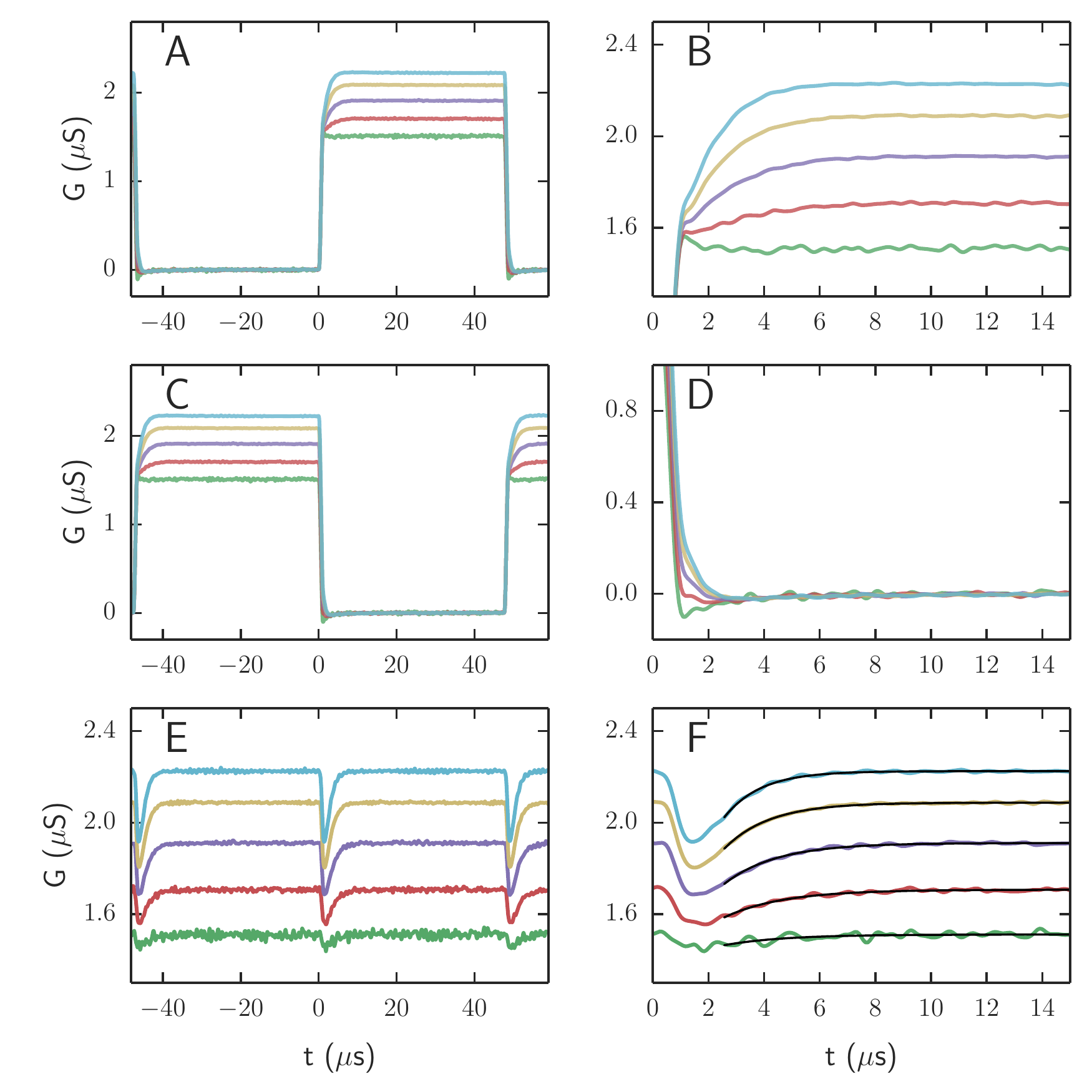}
  \caption{Response of sample to square wave with $V_{low} = 0$~mV and $V_{high} = $~1~mV (green), 2~mV (red), 3~mV (purple), 4~mV (yellow) and 5~mV (cyan). (A) and (B) show the response to the voltage turning on. (C) and (D) are shifted by half a period, such that (D) shows the ``turn-off" to 0~mV bias. (E) and (F) are obtained by adding the original and time-shifted signals, in order to cancel any voltage transient (due to, for example, LCR resonances in the wiring). The black lines in (F) are exponential fits to each curve. }
\end{figure}

Figure S1 shows example time traces at $\nu = 5/2$. Panel A shows the raw response of the sample to a square wave voltage, and panel B shows a zoomed in version showing increasing conductance after the voltage is turned on. In order to correct for electrical (non-thermal) transients, such as the slight decrease in the green curve in panel B, we shift the traces by half a period to obtain C and D and subtract them from the originals. Since the voltage bias is now returning to zero, there is no conductance (thermally sensitive) transient in panel D. By subtracting the original and time-shifted signals, we obtain the clean conductance shown in E and F. We then fit an exponential and extract the time constant $\tau$.

\subsection {Measurement by thermal relaxation}

In the main text, results were obtained by measuring the time constant for the temperature to increase in response to turning on Joule heating. In this section we present an alternative measurement of thermal relaxation when Joule heating is turned off. Again, conductance is used as the thermometer, which means that a small bias must still be applied during the ``heater off" time. Since that bias voltage must be kept small, the achievable signal-to-noise ratio (SNR) is limited. The data appear to saturate at $\tau=2 \, {\rm \mu s}$ since a relatively strong heater bias ($V_{high}$ = 5~mV) was used in order to obtain a discernible signal. As a result, the measurement reflects thermal relaxation from a temperature well above the final electron temperature, which is faster than would be measured for the final few millikelvin where $T_e \simeq T_{ph}$. Nonetheless, the discrepancy between the two data sets is at most 20\%.

\begin{figure}
  \centering
      \includegraphics[width=\columnwidth]{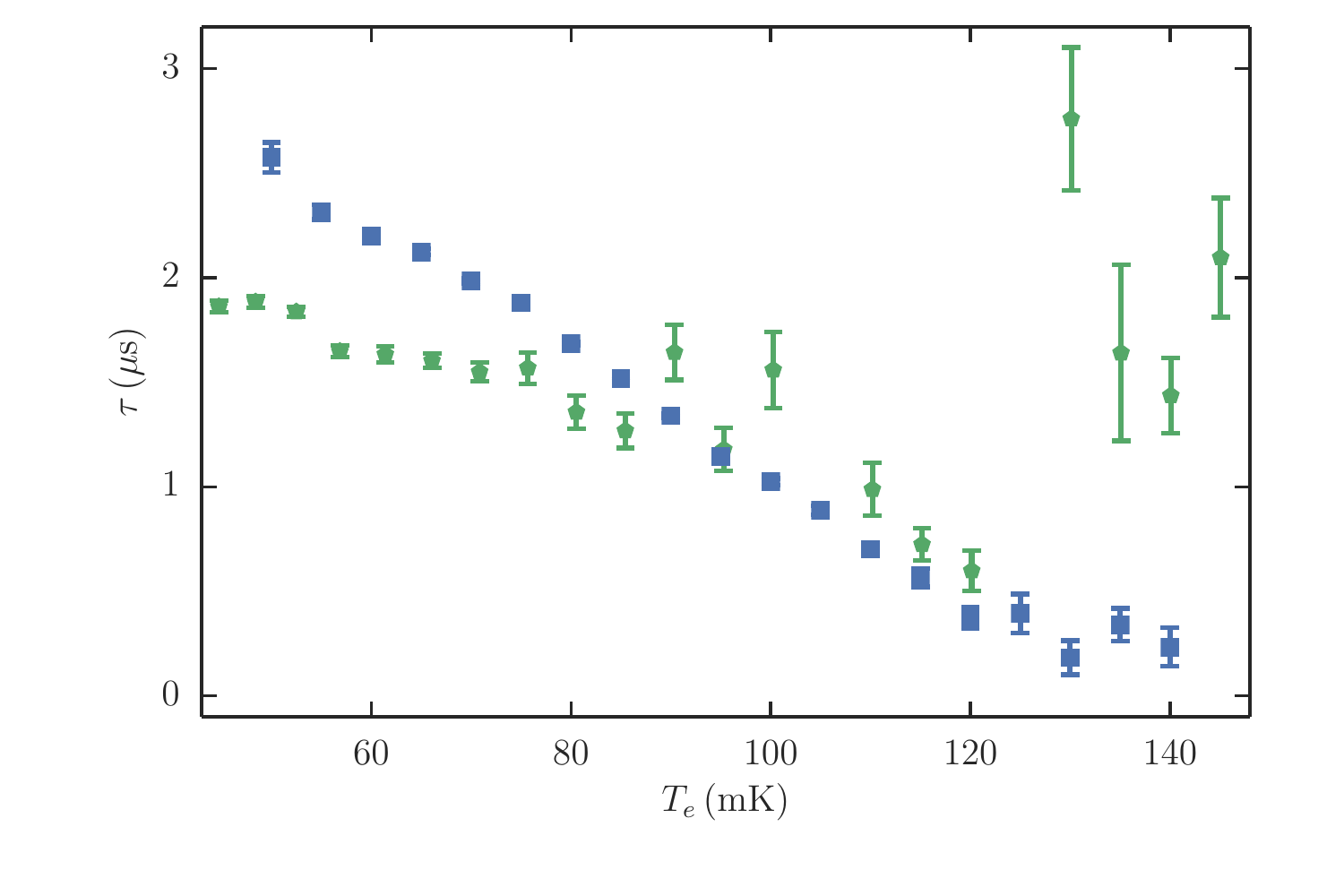}
  \caption{Comparison of time constants measured for heating (blue) and cooling using 1~mV probe bias voltage (green). Temperature values are the final electron temperature reached. Time constants for cooling deviate at low temperature since they measure the full relaxation from a temperature tens of mK higher than the final temperature. In contrast, the warming data only uses the final 10~mK or less, which is possible due to the higher SNR afforded by the larger bias used during the measurement.}
\end{figure}  

\section {Example of thermal conductance measurement}

\begin{figure}
  \centering
      \includegraphics[width=\columnwidth]{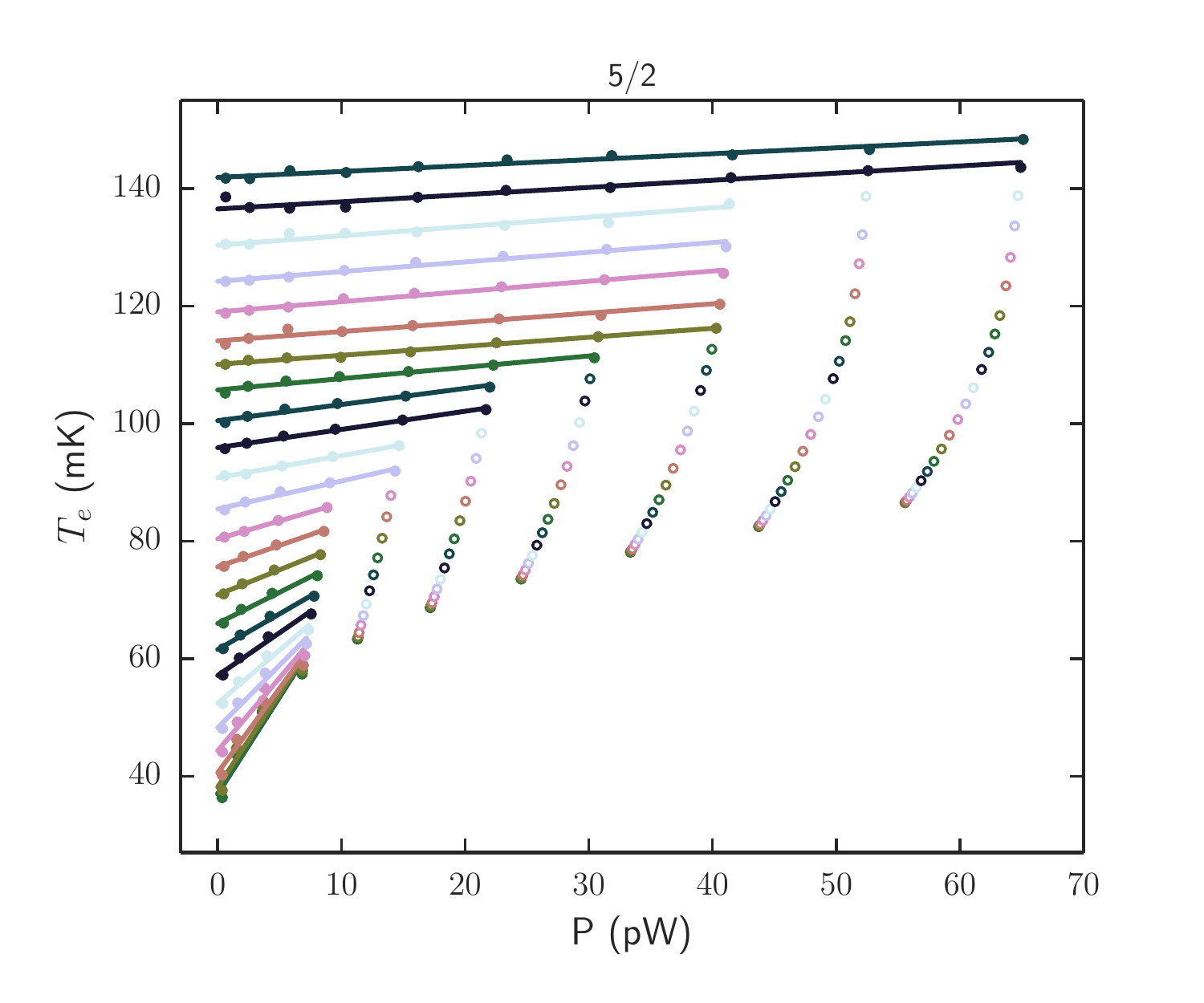}
  \caption{Electron temperature vs. Joule heating power at $\nu = 5/2$. Each data series corresponds to a different mixing chamber temperature (from 20 to 140~mK, in 5~mK increments). Voltage biases at each temperature were 0.5 to 5~mV in 0.5~mV increments. The linear fits were determined using  either the first 4 data points, or those that result in a temperature change of less than 7~mK. Filled markers indicate data points used in the fit, while empty markers were omitted}
\end{figure}  

Temperature vs. power data were extracted from the same datasets used to find $\tau$ (for example, the data shown in supplementary Fig.~1). The final conductance $G$ \textit{after} the transient was determined for each $T_{ph}$ and $V_{high}$ pair. A cubic spline interpolation to $G$ vs $T_e$ at the lowest $V_{high}$ was then used to map $G$ to $T_e$. These interpolations are shown as the solid curves in Fig.~3A of the main text. These were then used to determine $T_e (P, T_{ph})$, with ${P = GV_{high}^2}$. Supplementary Fig.~3 shows the resulting $T_e$ vs $P$ at $\nu = 5/2$ for $T_{ph}$ from 20 to 140~mK, with linear fits to the points where $T_e$ is only slightly higher than $T_{ph}$. The inverse of the slope of the fit gives $\kappa$, at $T_e$ given by the average temperatue of those points used in the fit.

\newpage  
\subsection {Geometric correction factor for the Corbino geometry}

The temperature change due to self-heating in the device is non-uniform, due to the non-constant current density in the Corbino geometry. More power is dissipated per unit area close to the center contact than at the outer edge, resulting in a higher temperature. We directly measure the apparent electron-phonon thermal conductivity, given by

\begin{equation}
\kappa_{measured} = \frac{P}{A} \frac{dR/dT}{\Delta R},
\label{equation:kappa_meas}
\end{equation} 

where $P$ is the total power dissipated in the 2DEG, $A$ is its area and $R$ is the resistance between the two contacts. In this section we calculate a geometric correction factor relating $\kappa_{measured}$ to the microscopic $\kappa$, 

\begin{equation}
\kappa = \frac{dP}{dA \Delta T},
\label{equation:kappa_actual}
\end{equation} 

where $dP$ is the power dissipated into a small area $dA$ chosen such that $\Delta T$ is uniform within the area $dA$. 

Consider the Corbino device to be made up of a series of concentric rings of radius $r$ and width $dr$. We define the inner radius as $r_1$, and the outer radius as $r_2$. For mathematical convenience, we work with resistance and resistivity, defined as $R \equiv 1/G_{rr}$ and $\rho \equiv 1/\sigma_{rr}$. Note that these are different from the longitudinal resistance and resistivity that would be measured in the Hall geometry.

The circumference of each ring is $2\pi r$, its area is $dA = 2\pi r dr$, and its contribution to the total resistance is $dR  = \rho (r) dr/(2\pi r)$. The total resistance of the Corbino is then found by integrating the contribution of each ring

\begin{equation}
R = \int_{r_1}^{r_2} { \frac{\rho(r) dr}{2\pi r}} 
\label{equation:R}
\end{equation}

For a small heating power,  $T(r)$ will only be higher than  $T_0$ by a small amount, which we denote by $\Delta T(r)$. We can then apply the first order Taylor expansion in resistivity, $\rho(T) \simeq \rho(T_0) + \frac{d\rho}{dT} \Delta T$, and the total resistance is then

\begin{equation}
R = R_0 + \frac{d\rho}{dT} \int_{r_1}^{r_2} { \frac{\Delta T(r) dr}{2\pi r}}, 
\label{equation:R}
\end{equation}

where $R_0 = R(T_0)$. For a small heating current, $i$, the power dissipated into each ring is given by

\begin{equation}
\delta P(r) = i^2 dR = \frac{i^2 \rho(r) dr}{2\pi r}.
\end{equation}

We can now find $\Delta T (r)$ using equation~2 as follows:

\begin{equation}
\Delta T(r) = \frac{1}{\kappa} \times \frac{i^2 \rho(r) dr}{(2 \pi r)^2 dr} \simeq \frac{1}{\kappa} \times \frac{i^2 \rho_0}{(2 \pi r)^2}, 
\label{equation:T}
\end{equation}

where we have made the simplification $\rho(r) \simeq \rho_0$, which is justified for $\Delta T \ll T_0$.
 
Substituting this into equation~\ref{equation:R}, we obtain

\begin{equation}
\Delta R \equiv R- R_0 = \frac{i^2 \rho_0}{(2\pi)^3\kappa} \frac{d\rho}{dT} \int_{r_1}^{r_2} \frac{dr}{r^3},
\label{equation:R2}
\end{equation} 

and

\begin{equation}
\kappa = \frac{-i^2 \rho_0}{(2\pi)^3\Delta R} \frac{d\rho}{dT} \left(\frac{1}{2}\right)\left(\frac{1}{r_2^2} - \frac{1}{r_1^2}\right)
\label{equation:kappa1}
\end{equation}  

Now, using the relations $\rho_0 = \frac{2\pi R_0}{\log(r_2/r_1)}$ and $\frac{d\rho}{dT} = \frac{2\pi}{\log(r_2/r_1)}\frac{dR}{dT}$, we can rewrite this in terms of the measured quantities $R_0$ and $dR/dT$,
 
\begin{equation}
\kappa = \frac{i^2 R_0}{4\pi \left(\log\left(r_2/r_1\right)\right)^2} \frac{dR/dT}{\Delta R} \left(\frac{1}{r_1^2} - \frac{1}{r_2^2}\right).
\label{equation:kappa2}
\end{equation} 

Finally, using the relations $P = i^2 R_0$ and $A = \pi (r_2^2 - r_1^2)$, we obtain

\begin{equation}
\kappa = \frac{P}{A} \frac{dR/dT}{\Delta R} \left(\frac{1}{2\log(r_2/r_1)} \left( \frac{1}{r_1^2} - \frac{1}{r_2^2}\right)\right)^2 
\label{equation:kappa3}
\end{equation} 
 
\begin{equation}
\kappa = \left(\frac{r_2^2 - r_1^2} {2 r_2 r_1 \log\left(r_2/r_1\right) } \right)^2 \times \kappa_{measured}
\end{equation}

For the sample discussed in this letter, $r_1 = 0.25$~mm and $r_2 = 1$~mm, giving a correction factor of $\sim$1.83. For a large heating current, the previous assumption that $\rho(r) \approx \rho_0$ in equation~\ref{equation:T} breaks down, and the correction factor in this case would have to be calculated numerically. In our measurement of $\kappa$, we use only the data for $\Delta T < 7$~mK and the correction factor of 1.83.

\begin{figure*}[h]
  \centering
      \includegraphics[width=\textwidth]{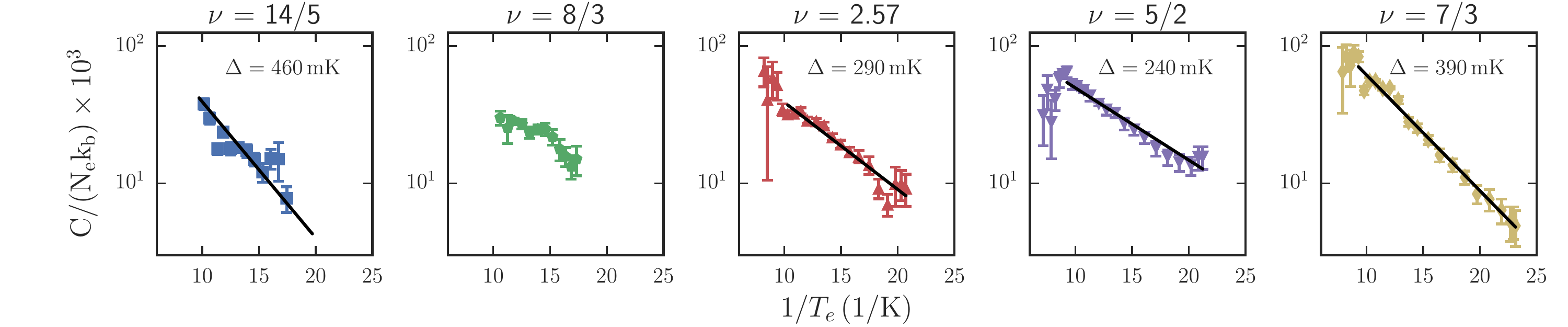}
  \caption{Arrhenius fits to the specific heat at each indicated filling factor, with a linear fit to the activated region (black lines). }  
\end{figure*}

\newpage 

\begin{figure*}[h]
  \centering
      \includegraphics[width=\textwidth]{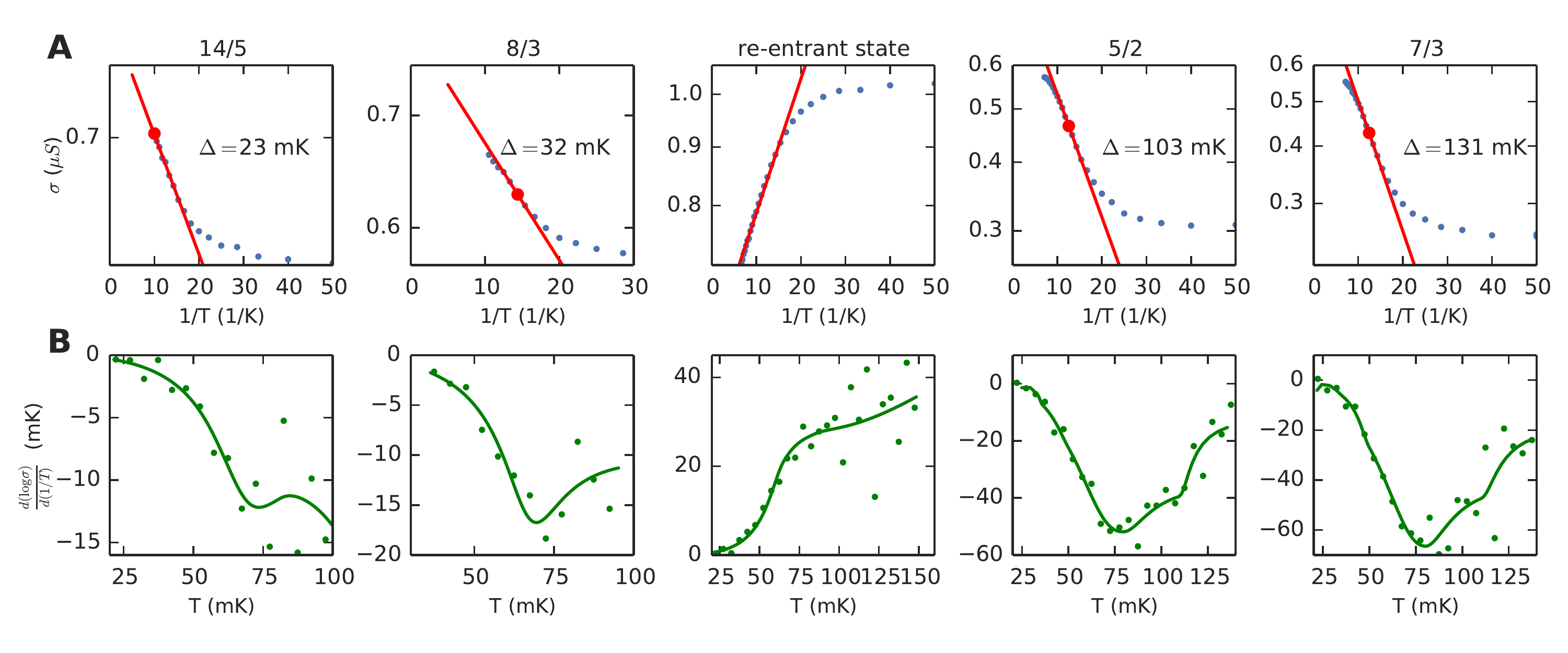}
  \caption{(A) Arrhenius fits to the conductivity at each indicated filling factor, with data in blue and linear fit to the activated region in red. The inflection point is marked by a red circle. (B) Corresponding plots of the Arrhenius slope vs. temperature, with data as points and a smoothed spline interpolation in green. The minima in (B) give $-\Delta/2$, and their locations in temperature give $T_i$. We find $T_i = 80$~mK at both $\nu = 5/2$ and $\nu = 7/3$, and $T_i \sim 100$~mK at $\nu = 14/5$. The ratio $\Delta_s/\Delta_i$ can than be estimated from figure~3 of d'Ambrumenil \textit{et. al.} [18]. Using this procedure we estimate saddle point gaps of 300~mK and 330~mK for $\nu = 5/2$ and $\nu = 7/3$, respectively. At $\nu = 14/5$, we have $\Delta/T_i \sim 0.25$, which would imply a saddle-point gap around 20 times the transport gap, roughly agreeing with our observed gap in specific heat. However, it would also require tiny saddle point widths, and interpretation of the $\nu = 14/5$ gap should be taken as purely speculative until a more detailed study can be performed.  }  
\end{figure*}

\clearpage  
\section {High integer filling factor data}
 \renewcommand\floatpagefraction{0.8}
 
For well-separated Landau levels ($ T << \hbar \omega_c$ and $\Gamma << \hbar \omega_c$, where $\Gamma$ is the LL broadening), the integral of the heat capacity between two integer filling factors, $\nu = i$ and $\nu = i+1$, is given by

\begin{equation}
\int_{B_{\nu=i+1}}^{B_{\nu=i}} c(T, B) dB = \frac{4\pi^2m^*}{3(i - 1/2)^2e \hbar} k_b T. 
\end{equation}

This result holds for a symmetric broadening function. Numerical integration of the results in figure~S6D from $\nu = 11$ to $\nu = 12$ gives a value of $4.4 \times 10^{-4} \, {\rm \left[k_b\cdot T\right]}$ per electron, while the theoretical result is $3.0 \times 10^{-4} {\rm \left[k_b\cdot T\right]}$ per electron. The discrepancy may be due to the onset of strongly interacting bubble and stripe phases.

\begin{figure}
    \centering
      \includegraphics[width=\columnwidth]{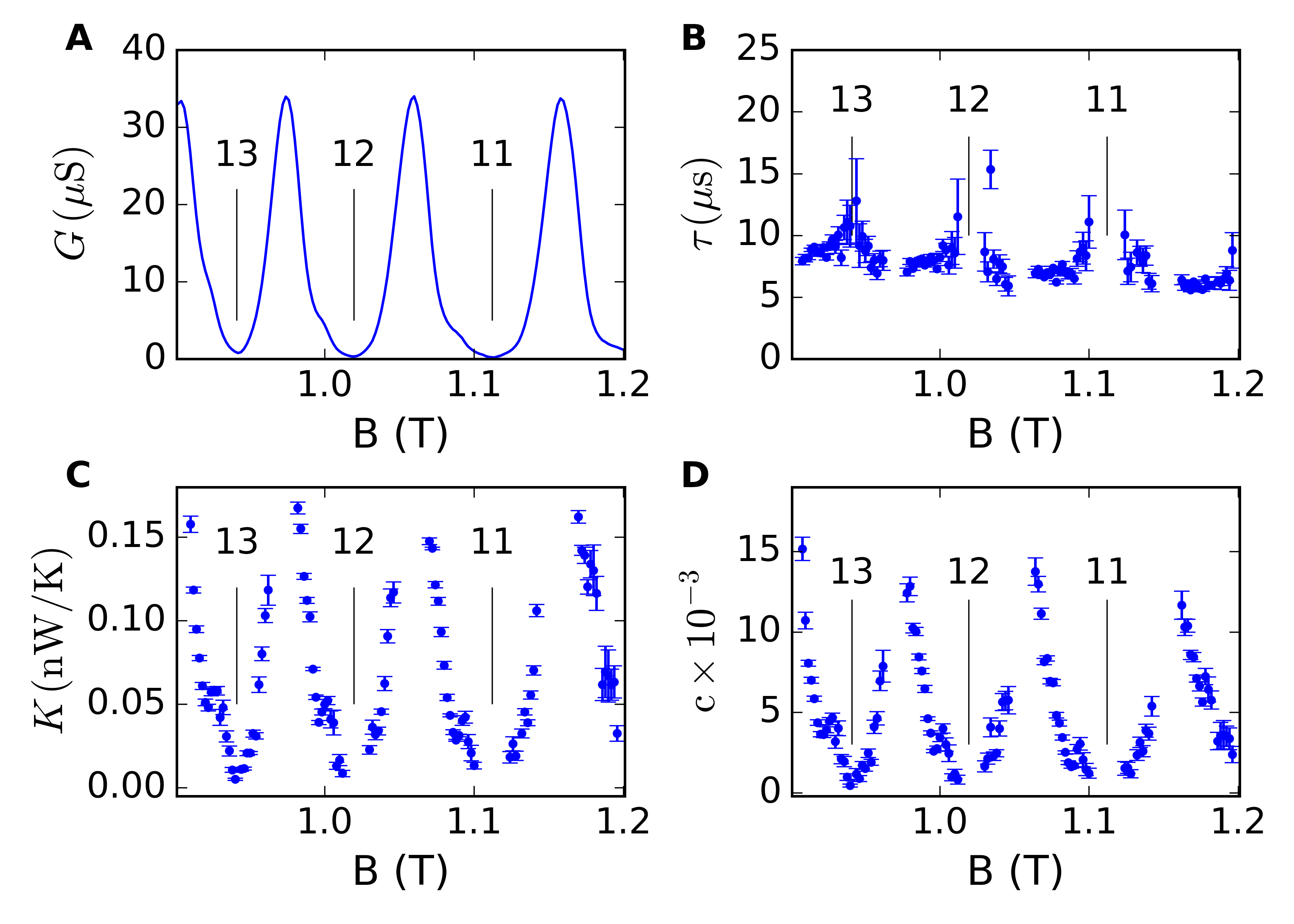}
  \caption{Example of data for high Landau levels at 40~mK. (a) Conductance trace, showing IQH minima at labeled filling factors as well as onset of bubble phases on the left flank of each minimum. (b)  Thermal relaxation times, measured during cooling. We find that $\tau$ has only a weak dependence on filling factor within this range of field, but is systematically longer than in the SLL data shown in the main text. (c) Thermal conductance, $K$, varies strongly with filling factor, roughly following the trend of $G$. This is reasonable, since both $K$ and $G$ are sensitive to the density of states near the Fermi energy. (d) Specific heat calculated from $\tau$ and $\kappa$. We find the expected oscillations of $c$ as the DOS near the fermi energy varies substantially with filling factor. }  
\end{figure}

\end{document}